# Spiking Photonic Neurons Based on Two-Section InP Quantum-Well Lasers Integrated on Silicon

**MENELAOS SKONTRANIS,[1,*] BENOIT CHARBONNIER,[2,†] OLIVIER GIRARD,[3,†] ADONIS BOGRIS,[4,†], CHARIS MESARITAKIS,[1,†]**

[1]*Department of Biomedical Engineering, University of West Attica, Agiou Spiridonos,12243, Egaleo, Athens, Greece*
[2] *CEA-Leti, Lyon, Auvergne-Rhône-Alpes, France*
[3] *Scintil Photonics, 69 rue Félix Esclangon 38000 Grenoble, France*
[4]*Department of Informatics & Computer Engineering, University of West Attica, Agiou Spiridonos, 12243, Egaleo, Athens, Greece*
[†]*M.S. conducted the experiments and analyzed the data, B.C. and O. G. fabricated the chip, A.B. and C.M. supervised the research*
*mskontranis@uniwa.gr

**In this work we experimentally investigate the spiking dynamics of two-section InP quantum-well lasers monolithically integrated on silicon. By appropriately tuning the electrical bias conditions, we realize multiple neuronal-like operating regimes, such as integrate-and-fire and resonate-and-fire, highlighting the device's versatility as a high-speed photonic neuron. A systematic investigation of laser design parameters, including cavity length and gain/saturable absorber ratio, elucidates their impact on spiking-related properties (such as pulse repetition frequency) and traces the operational parameter space that unlocks stable spiking. Finally, these findings pave the way toward scalable neuromorphic photonic integrated circuits, where low-loss silicon synapses coexist with versatile laser neurons.**

The rapid growth of data-intensive and artificial intelligence workloads is increasingly constrained by fundamental limitations of conventional computing architectures, such as memory–bandwidth bottlenecks, scaling challenges, rising power consumption, and increased thermal management demands [1]. These challenges have motivated the exploration of alternative computing paradigms spanning both architecture and materials. In this context, photonic neuromorphic computing has emerged as a promising solution, combining the intrinsic advantages of photonics—ultrafast dynamics, ultrabroad bandwidth, intrinsic parallelism, energy-efficient operation, and wavelength- and time- multiplexing—with the spike-based information processing principles of neuromorphic computing [2,3].

Within this framework, Photonic Integrated Circuits (PICs) have emerged as a key platform for compact and scalable photonic neuromorphic systems. In this context, recent efforts have focused on passive–active co-integration, enabling the integration of III–V active devices with low-loss silicon photonic circuitry and reducing the complexity of conventional hybrid integration approaches [4,5]. This approach is particularly attractive for neuromorphic architectures, where spiking laser neurons can be efficiently interconnected via low-loss silicon waveguide-based synapses. Beyond integration, optimizing the dynamical behavior of spiking light sources, such as two-section gain saturable absorber (SA) lasers, is critical for improving neuromorphic system performance. Parameters such as spike duration, directly affects the temporal resolution of spiking neural networks, when time-delay encoding strategies are assumed, whereas the neural refractory period is linked to processing latency of laser based neuromorphic schemes. Although these dynamics have been extensively studied in conventional III/V two-section laser sources [6,7], dynamics in monolithically integrated gain–SA lasers on silicon have not yet been investigated.

In this work, we present an experimental investigation of the spiking dynamics in III-V on silicon quantum-well two-section lasers. By independently tuning the gain current and the reverse bias applied to the SA ($V_{ABS}$), as well as varying key structural parameters, we establish clear correlations between device design, operating conditions and different dynamical spiking regimes. By highlighting the capabilities of these CMOS-compatible lasers we pave the way towards a holistic neuromorphic PICs, where versatile spiking neurons will coexist with low-loss silicon interconnects, acting as neural synapses.

To support this investigation, the PIC of Fig.1a was fabricated, comprising an array of multiple two-section gain–SA lasers, each uniquely defined by its cavity length (L) and the ratio of the SA length to L ($R_{ABS}$). Each laser had two independently biasing sections, designated as gain and SA sections, whereas Sagnac loops implemented the facets of the laser cavity offering reflectivities of 10% and 40%.

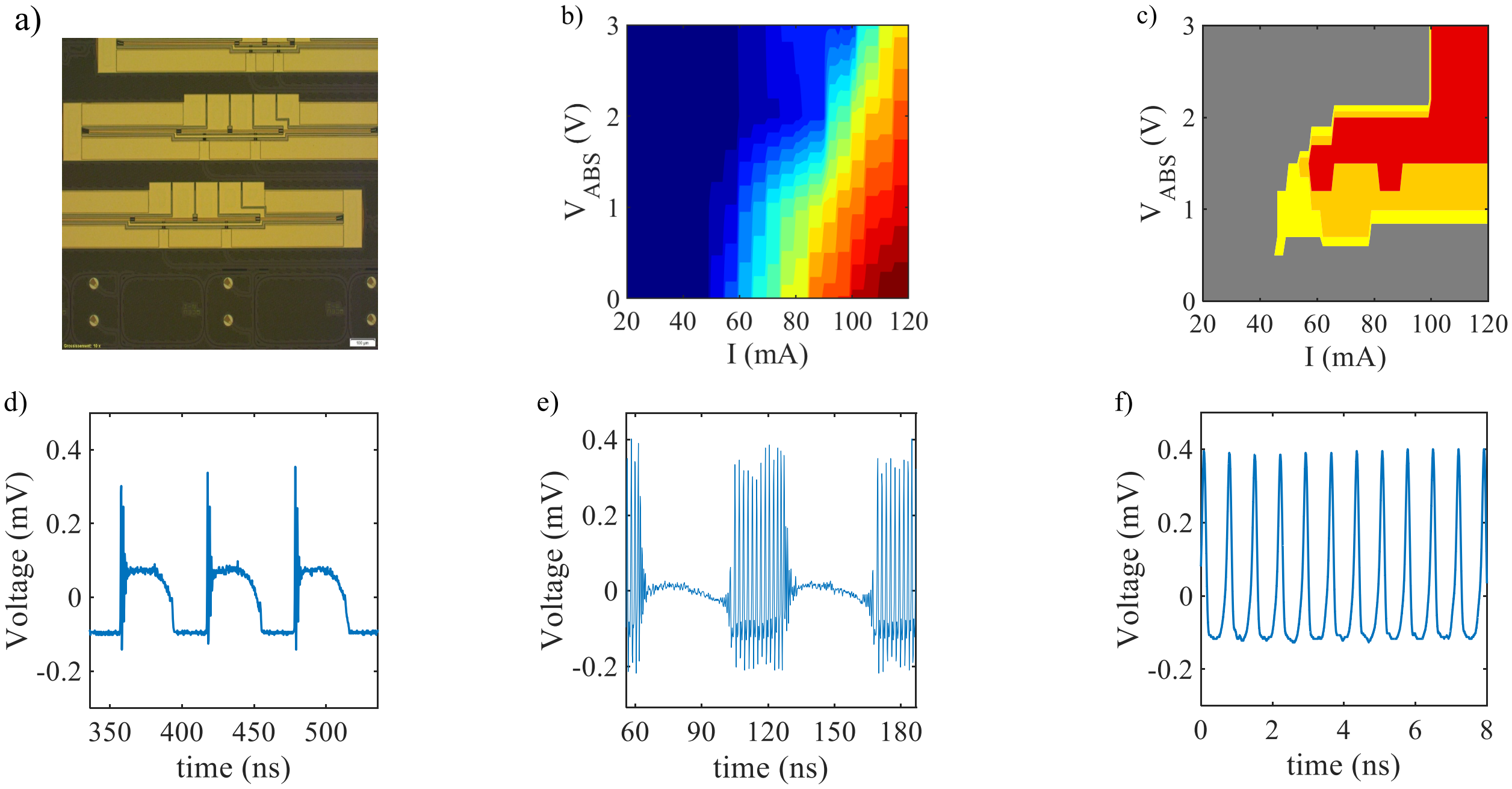

**Fig. 1** a) picture of the PIC. Characterization of the laser with L=800μm and $R_{ABS}$=3% b) P-I heatmap, c) map with the regimes of the laser. In the grey area, no pulsations were observed whereas in the yellow, orange and red area pulses corresponding to regime I, II and III are recorded. Time traces of pulses from regime d) I, e) II and f) III.

During characterization the PIC was mounted on a temperature-stabilized copper stage maintained at 25 °C via a thermo-electric cooler in a closed loop, while electrical pumping and SA biasing were applied using tungsten probes mounted on XYZ micro-positioners. Light was collected through 7° tilted as-cleaved single-mode fiber (SMF), inducing coupling losses of approximately 6-7dB. The laser's spiking time traces were monitored using a 10 GHz photodiode followed by an 8 GHz/80Gsample/s real-time oscilloscope. Electrical excitation to the laser neuron was generated through an 8GHz Arbitrary waveform generator.

Initial characterization consisted measuring of the optical power versus both the injection current and reverse voltage at the SA ($V_{ABS}$), for a laser with L = 800 μm and $R_{ABS}$ = 3% (Fig.1b). Lasing threshold is achieved at 42 mA, when $V_{ABS}$=0V with an external quantum efficiency of 24.6μW/mA. Increasing $V_{ABS}$ to 3V raises threshold to 60 mA and reduces efficiency to 13.1μW/mA due to increased unsaturated losses by the SA [6]. The relatively low quantum efficiency of the device is attributed mainly to the low reflectivity offered by the Sagnac loops.

Subsequently, the self-pulsation dynamics were investigated. The measurements revealed three distinct operating regimes, denoted as Regimes I, II, and III, highlighted in yellow, orange, and red in Fig.1c respectively. These regimes were achieved for different biasing conditions and mainly differed in terms of pulse repetition frequency (PRF), full width at half maximum (FWHM), and peak power (PP) of the generated spikes. Regime I can be achieved at low injection current and low $V_{ABS}$ and is characterized by temporally broad pulses with FWHM ranging from 14 to 38 ns (Fig. 1d). As injection current and $V_{ABS}$ increase, the laser transitions into Regime II, exhibiting burst dynamics where broad pulses and sharp spikes (FWHM ≈ 120-140 ps) coexist (Fig. 1e). At higher injection currents (> 80 mA) and $V_{ABS}$ (> 1 V), the laser enters Regime III, where emission is dominated by stable sub-nanosecond pulses (Fig. 1f). Although burst spiking, observed in regime II is relevant to biological neural operation, here we focus mainly on Regime I and III as these excitable regimes are particularly relevant for neuromorphic photonics. In particular, they allow the realization of the two types of excitable neurons: integrate-and-fire and resonate-and-fire neurons. The difference between these computational primitives can be traced on the dependence of the PRF and spike amplitude on bias current [8,9].

In this context, Regime I pulses are linked to resonate-and-fire behavior as they are characterized by a nearly constant PRF of 16 MHz (Fig.2a) independently of injection strength, whereas a constant spike amplitude increase with injection current can be monitored (Fig.2b) [8,9]. The physical mechanism for this behavior is linked to Canard spikes, attributed to optothermal dynamics in two section SA semiconductor lasers [10]. On the other hand, Regime III pulses exhibit an integrate-and-fire like behavior as the PRF shows a strong and continuous increase with injection current, reaching frequencies as high as 1.8 GHz (Fig. 2c) whereas the spike amplitude rapidly saturates shortly after threshold (the onset of pulsations) (Fig. 2d). In this case, the generation of pulses is attributed to gain–SA and coupled dynamics, commonly described as Q-switching operation, where the interplay between carrier buildup in the gain section and intensity-dependent absorption in the SA governs pulse formation and timing [6,11]. Interestingly both Regimes I and

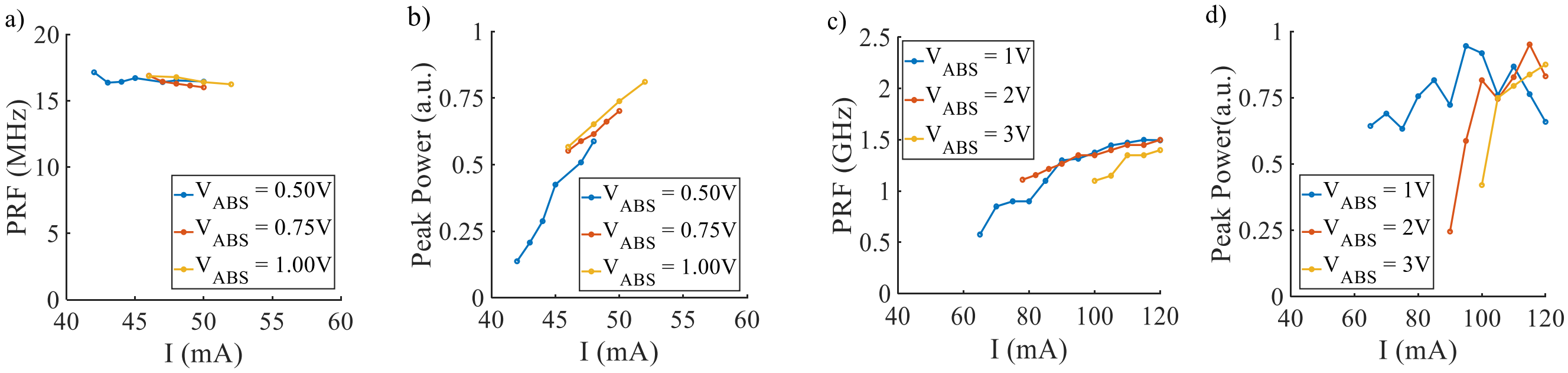

**Fig. 2** a) PRF and b) PP of Regime I pulses for various bias and $V_{ABS}$. c) PRF and d) peak power of Regime III pulses for various bias ar

III are present in the same physical device, triggered only through a minor change in the DC biasing. Therefore, the same laser can demonstrate both resonate-and-fire and integrate-and-fire neural like behavior.

Aiming to further validate the neural isomorphism of Regime III, the three key neuronal attributes associated with this dynamic state, have been investigated: The existence of a spiking threshold, temporal integration, and a refractory period. To this end, we biased the laser at I=71.24 mA and $V_{ABS}$=-2.2 V, placing its operating point near the boundary of Regime III (see Fig.1c). Excitability was triggered by applying external positive voltage pulses to the ABS with a FWHM of 0.5 ns and variable amplitude. As shown in Fig. 3a, a single 0.8 V pulse reliably elicits an optical spike, whereas a 0.3 V pulse (subthreshold) fails to do so, confirming the existence of a threshold. However, when two sub-threshold pulses are applied with a temporal separation of 0.4 ns, their combined effect is sufficient to trigger a spike, demonstrating temporal integration. Increasing the pulse separation to 1.1ns leads to no spike generation (Fig.3b), indicating that the accumulated excitation decays over time. Finally, the refractory behavior is verified by applying two suprathreshold pulses. For a pulse separation of 0.5 ns, only a single spike is observed (Fig. 3c), while increasing the delay to 0.8 ns results in two distinct spikes (Fig. 3d), demonstrating recovery of the excitable state after a finite refractory period. This last claim is also aligned with the maximum PRF associate with this regime and device that dictates a minimum resting period of 0.55ns < 0.8ns.

Furthermore, we investigated the impact of two laser design parameters in terms of expanding the bias conditions that allow integrate and fire (regime III). These parameters are L and $R_{ABS}$. To this end, operational regime maps were constructed versus bias conditions for nine different lasers, corresponding to L=400, 600 and 800 μm and $R_{ABS}$=3, 5 and 8% (for L=600 μm we used $R_{ABS}$=10% instead of 8%). Our results indicate that for $R_{ABS}$=3% and 5%, all three self-pulsation regimes are present (Fig.4a-f), whereas for $R_{ABS} >$ 8%, only Regime I is observed (Fig.4g-h). An exception occurs for L=800 μm and $R_{ABS}$=8%, where Regimes II and III appear over a very narrow range of bias conditions (Fig.4i). Increasing $R_{ABS}$ reduces the area of Regime III and shifts it toward higher bias currents. This behavior arises since longer SA lengths increase the corresponding saturation energy, requiring higher intracavity energy to bleach the SA and initiate pulse generation. In general, for implementing fast integrate-and-fire neurons, $R_{ABS}$ should not exceed 5%.

Regarding cavity length, increasing L expands the range of biasing conditions in which regime III occurs and enables higher PRFs. As shown in Fig. 4, the maximum PRF, indicated with an X in Fig.4, increases with L, likely because for higher L, a larger number of photons is produced per round trip, allowing a faster accumulation of energy in the absorber which enables faster self-pulsations. This behavior was also observed in previous works [12]. Interestingly, the highest PRF is not obtained for the lowest $R_{ABS}$=3%, but rather for $R_{ABS}$ = 5%. This counterintuitive result is probably attributed to the onset of Regime III at lower $V_{ABS}$ (1 V for $R_{ABS}$ = 5% compared to 1.5 V for $R_{ABS}$ = 3%), which enables operation under reduced effective intracavity losses and consequently supports higher maximum repetition rates.

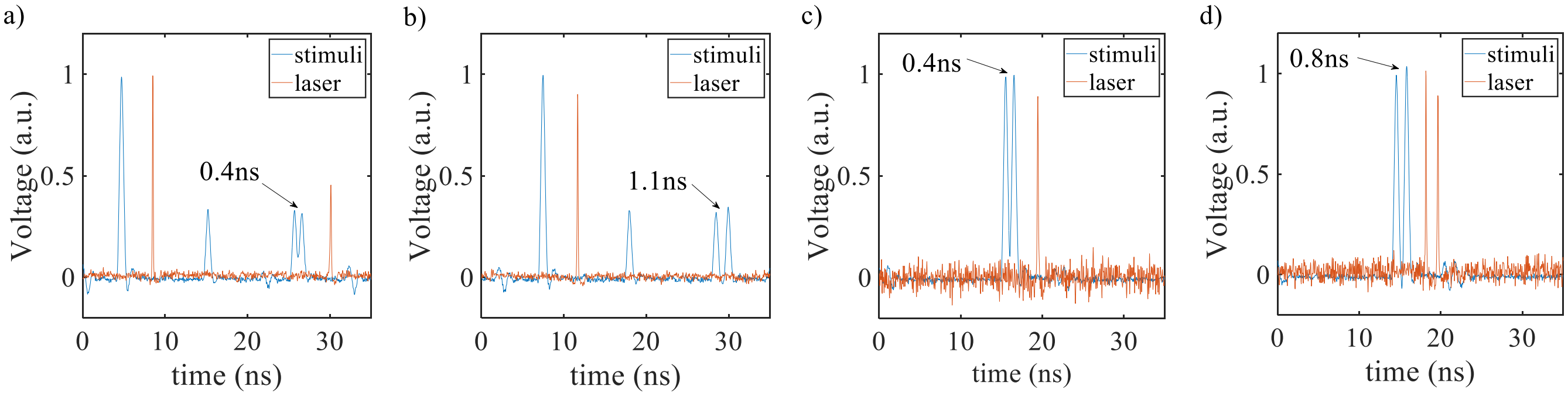

**Fig. 3** Dynamical response of the L=800μm and $R_{ABS}$=3% laser for electrical pulses of 0.8V, followed by a sub-threshold pulse of 0.3V and a pair of subthreshold pulses of 0.3V. The time interval between the pair of subthreshold pulses is 0.4ns (a) and 1.1ns (b). Dynamical response of the L=800μm and $R_{ABS}$=3% laser for a pair of suprathreshold pulses of 0.8V. The time interval between the pair of suprathreshold pulses is 0.4ns (c) and 0.8ns (d). All perturbation pulses have a FWHM of 0.5ns

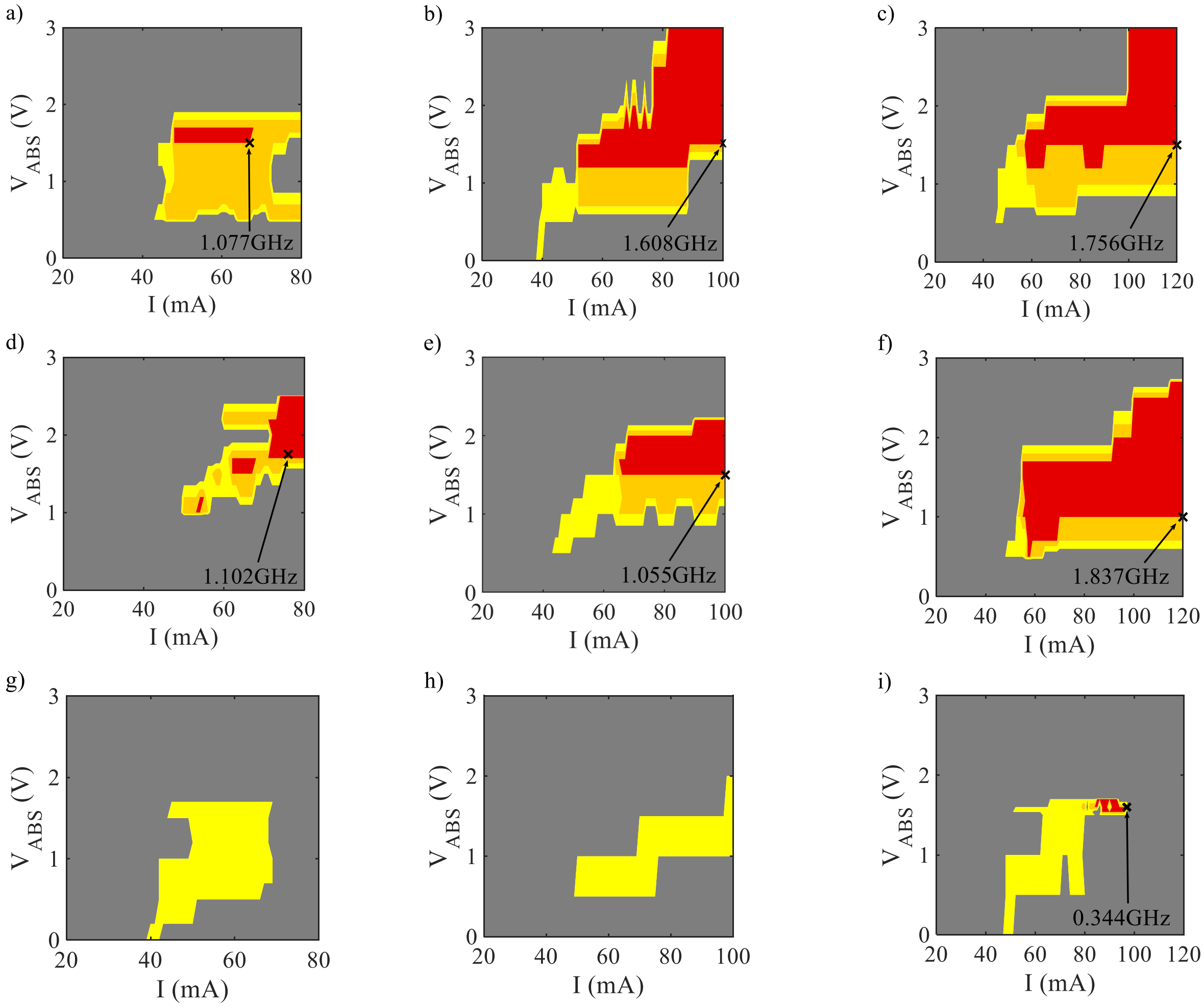

**Fig. 4** Map regimes for $R_{ABS}$=3% and L=400, 600 and 800µm (a-c), for $R_{ABS}$=5% and L=400, 600 and 800µm (d-f) and for $R_{ABS}$=8(10)% and L=400 and 800 (600)µm (g-i). X marks the bias condition with the highest detected PRF in regime III.

In conclusion, we demonstrated that III/V two-section quantum-well lasers fabricated on silicon can provide both resonate-and-fire and integrate-and-fire dynamics, with distinct time scales arising from different underlying mechanisms (canard- and Q-switching), via varying the DC bias conditions. Due to its significant neurocomputational properties, we focused on the integrate-and-fire regime, confirming its neural isomorphism and optimizing its performance for two key structural parameters. Laser length (L) and gain/absorber ratio ($R_{ABS}$) govern this regime, where increasing L enhances PRF, enabling faster operation.

**Funding.** This work is supported by EU HORIZON project PROMETHEUS (ID: 101070195).

**Disclosures.** The authors declare no conflicts of interest.

**Data availability.** Data underlying the results presented in this paper are not publicly available at this time but may be obtained from the authors upon reasonable request.

## References

1. Mitchell Waldrop, Nature **530**, 144 (2016).
2. D. Marković, A. Mizrahi, D. Querlioz, et al, Nat Rev Phys **2**, 499 (2020).
3. B. J. Shastri, A. N. Tait, T. Ferreira de Lima, W. H. P. Pernice, H. Bhaskaran, C. D. Wright, and P. R. Prucnal, Nat. Photonics **15**, 102 (2021).
4. S. Shekhar, W. Bogaerts, L. Chrostowski, J. E. Bowers, M. Hochberg, R. Soref, and B. J. Shastri, Nat Commun **15**, 751 (2024).
5. N. Farmakidis, B. Dong, and H. Bhaskaran, Nat Rev Electr Eng **1**, 358 (2024).
6. G. J. Spühler, R. Paschotta, R. Fluck, B. Braun, M. Moser, G. Zhang, E. Gini, and U. Keller, J. Opt. Soc. Am. B **16**, 376 (1999).
7. H. Uenohara, R. Takahashi, Y. Kawamura, and H. Iwamura, IEEE J. Quantum Electron. **32**, 873 (1996).
8. P. R. Prucnal, B. J. Shastri, T. Ferreira De Lima, M. A. Nahmias, and A. N. Tait, Adv. Opt. Photon. **8**, 228 (2016).
9. E. M. Izhikevich, Computational Neuroscience (MIT press, 2007).
10. A. Tierno, N. Radwell, and T. Ackemann, Phys. Rev. A **84**, 043828 (2011).
11. D. Rachinskii, A. Vladimirov, U. Bandelow, B. Hüttl, and R. Kaiser, J. Opt. Soc. Am. B **23**, 663 (2006).
12. V. Z. Tronciu, M. Yamada, T. Ohno, S. Ito, T. Kawakami, and M. Taneya, IEEE J. Quantum Electron. **39**, 1509 (2003).

## References

[1] Waldrop, M. M. (2016). More than Moore. *Nature*, *530*(7589), 144-148.

[2] Marković, D., Mizrahi, A., Querlioz, D., & Grollier, J. (2020). Physics for neuromorphic computing. *Nature Reviews Physics*, *2*(9), 499-510.

[3] Shastri, B. J., Tait, A. N., Ferreira de Lima, T., Pernice, W. H., Bhaskaran, H., Wright, C. D., & Prucnal, P. R. (2021). Photonics for artificial intelligence and neuromorphic computing. *Nature Photonics*, *15*(2), 102-114.

[4] Shekhar, S., Bogaerts, W., Chrostowski, L., Bowers, J. E., Hochberg, M., Soref, R., & Shastri, B. J. (2024). Roadmapping the next generation of silicon photonics. *Nature Communications*, *15*(1), 751.

[5] Farmakidis, N., Dong, B., & Bhaskaran, H. (2024). Integrated photonic neuromorphic computing: opportunities and challenges. *Nature Reviews Electrical Engineering*, *1*(6), 358-373.

[6] Spühler, G. J., Paschotta, R., Fluck, R., Braun, B., Moser, M., Zhang, G., ... & Keller, U. (1999). Experimentally confirmed design guidelines for passively Q-switched microchip lasers using semiconductor saturable absorbers. *Journal of the Optical Society of America B*, *16*(3), 376-388.

[7] Uenohara, H., Takahashi, R., Kawamura, Y., & Iwamura, H. (2002). Static and dynamic response of multiple-quantum-well voltage-controlled bistable laser diodes. *IEEE journal of quantum electronics*, *32*(5), 873-883.

[8] Prucnal, P. R., Shastri, B. J., Ferreira de Lima, T., Nahmias, M. A., & Tait, A. N. (2016). Recent progress in semiconductor excitable lasers for photonic spike processing. *Advances in Optics and Photonics*, *8*(2), 228-299.

[9] Izhikevich, E. M. (2007). *Dynamical systems in neuroscience*. MIT press.

[10] Tierno, A., Radwell, N., & Ackemann, T. (2011). Low-frequency self-pulsing in single-section quantum-dot laser diodes and its relation to optothermal pulsations. *Physical Review A—Atomic, Molecular, and Optical Physics*, *84*(4), 043828.

[11] 11. D. Rachinskii, A. Vladimirov, U. Bandelow, B. Hüttl, and R. Kaiser, J. Opt. Soc. Am. B 23, 663 (2006).

[12] Tronciu, V. Z., Yamada, M., Ohno, T., Ito, S., Kawakami, T., & Taneya, M. (2003). Self-pulsation in an InGaN laser-theory and experiment. *IEEE journal of quantum electronics*, *39*(12), 1509-1514.